\documentclass[aps,prl,twocolumn,groupedaddress]{revtex4-1}
\usepackage{graphicx}
\usepackage{graphicx}
\usepackage{dcolumn}
\usepackage{bm}
\usepackage{color}
\usepackage{amsmath}

\usepackage{array}

\begin{document}

\title{Crossover from Hydrodynamics to the Kinetic Regime in Confined Nanoflows}
\newcommand {\BU}{Department of Mechanical Engineering, Boston University, Boston, Massachusetts 02215, USA.}


\author{C. Lissandrello}
\affiliation{\BU}
\author{V. Yakhot}
\affiliation{\BU}
\author{K. L. Ekinci}
\email[e-mail:] {ekinci@bu.edu}
\affiliation{\BU}


\date{\today}


\begin{abstract}

We present an experimental study of a confined nanoflow, which is generated  by a sphere oscillating in the proximity of a flat solid wall in a simple fluid. Varying the oscillation frequency, the confining length scale and the fluid mean free path over a broad range provides a detailed map of the flow. We use this experimental map to construct a scaling function, which describes the nanoflow in the entire parameter space, including both the hydrodynamic and the kinetic regimes. Our scaling function unifies previous theories based on the slip boundary condition and the effective viscosity.

\end{abstract}


\maketitle

In micron and nanometer scale flows \cite{Karniadakis,Tabeling},  the characteristic dynamic length scale ${\cal L}$ of the flow approaches and is even exceeded by the mean free path of the fluid $\lambda$. This limit is clearly beyond the  applicability of the Navier-Stokes equations, requiring a rigorous treatment using  kinetic theory. A less rigorous but widely used approach to describe these small scale flows is to extend the Newtonian description by imposing a slip boundary condition on solid walls. This approach is justified as follows. Derivation of the Navier-Stokes equations from kinetic theory  results in the appearance of a Knudsen  layer of  thickness $\lambda$ near the wall \cite{Lifshitz_Kinetics}. Because  a  fluid element of linear dimension  $\sim\lambda$ is treated as a mathematical point in the hydrodynamic approximation, the velocity at the wall  becomes ${u_w} \approx \lambda {\left. {{{du} \over {dz}}} \right|_{z = 0}}$, with $u$ being the hydrodynamic velocity (assumed parallel to the wall) and ${\bf{\hat z}}$ being the wall normal. Thus, the slip length $b$, where $b \sim \lambda$, is applied as a convenient empirical parameter to extend the Navier-Stokes equations into the kinetic regime. As required by macroscopic hydrodynamics, $b$ becomes negligible when the Knudsen number, ${\rm{Kn}} \equiv {\lambda  \over {\cal L}}$, is small, i.e., ${\rm{Kn}} \ll 1$.

The above approach comes with some problems. To describe some gas flows, for instance, unphysical slip lengths, $ b\gg \lambda$, may be required. To alleviate this problem, one can assume specular reflections of the gas molecules from the wall. Yet, experiments show that this assumption is not very accurate for heavier gases and untreated surfaces \cite{Arkilic,Trott}. Worse is the problem when the Navier-Stokes solution (with the slip boundary condition) fails to converge with the prediction of  the kinetic theory. A  good example to the point is oscillating nanoflows \cite{Ekinci_Nanofluidics, Svitelskiy, TJ_cantilever}. Efforts to describe oscillating nanoflows using the Navier-Stokes equations in conjunction with a slip length agree with experiments only in a  range of relevant parameters \cite{Bhiladvala_Molecular}. A proper kinetic treatment of the problem \cite{Yakhot_Colosqui} shows why: the finite relaxation time $\tau$ of the fluid modifies the physics of the flow, resulting in the ``telegrapher's equation'', which is  substantially different from the Navier-Stokes equations.


\begin{figure*}
\includegraphics[width=7in]{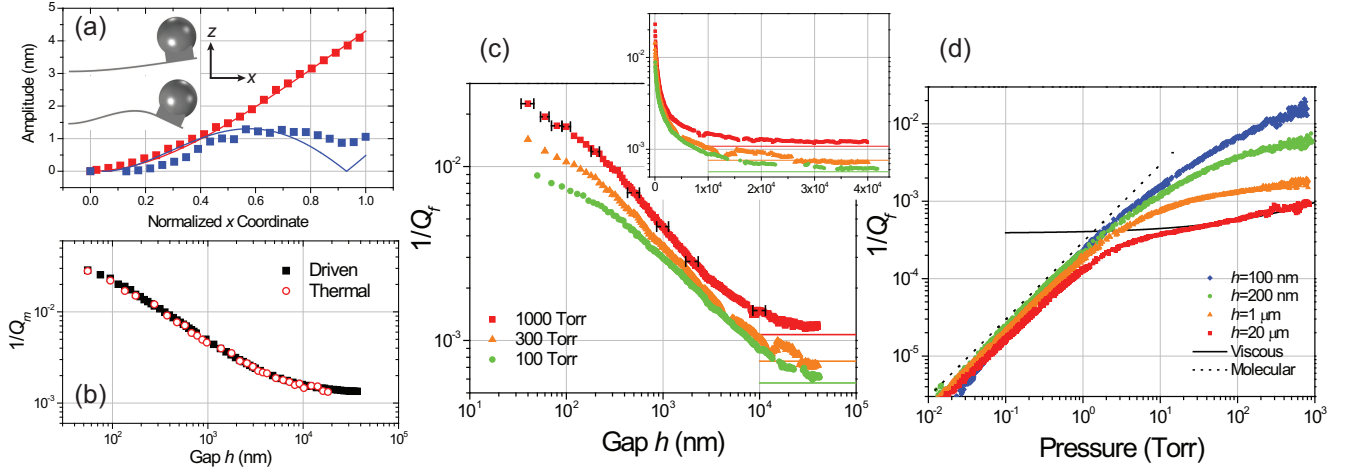}
\caption{\label{fig_asymptote} (a) Measured (symbols) and simulated (inset and solid lines) mechanical modes of the sphere-cantilever device. In the first harmonic mode (blue), FEM simulations suggest that the node appears at the position where the sphere is attached to the cantilever and that the sphere undergoes small rotational oscillations about an axis (parallel to the $y$-axis) through the node. (b) Measured dimensionless dissipation $1/Q_m$ as a function of gap $h$, obtained from the driven frequency response, and thermal oscillations of the device. (c) Dimensionless fluidic dissipation $1/Q_f$ as a function of gap $h$ at fixed pressures $p$. Solid line segments show asymptotic values $1/Q_{f\infty}$ from viscous theory. Inset is a semi-logarithmic plot of the same data. The $h$ error bars for $h\le200$ nm are due to roughness and contact uncertainty. Otherwise, the error comes from the linear stage. The error bars in $1/Q_f$ are smaller than the symbol sizes. (d)  $1/Q_f$ measured as a function of pressure $p$ with the gap fixed. Solid line is from viscous theory and the dotted line is from molecular theory.  All data in this figure are obtained from device C1.}
\end{figure*}


In this manuscript, we turn our attention to  nanometer scale  confined flows in the limit $h\lesssim \lambda$, where $h$ is the confining length scale. So far, a group of researchers have  extended Reynolds' hydrodynamic formulation \cite{Reynolds_1886,Vinogradova_1995} to  small scales  by imposing the slip boundary condition \cite{Ducker_2010, Ducker_2011, Bhushan, Chevrier_2011,Bhiladvala_Knudsen} --- as described above. Others, coming from kinetic theory,  have developed the concept of  the effective viscosity, which typically depends upon a properly defined Knudsen number  \cite{Bao_Squeeze,Veijola}. There is no question that both approaches must agree for the  same flow parameter space.  Here, we present an  experimental study of nanometer scale  confined flows covering a broad range of parameters --- including gap $h$, pressure $p$, and frequency $\omega\over2\pi$ --- along with a scaling theory. Our scaling function describes the physical behavior of the flow in the entire parameter space, capturing the transition from  hydrodynamics to the kinetic regime accurately.

We  study the oscillatory hydrodynamic response of a  sphere in the proximity of a solid surface. Our experimental device is a micron-scale silica sphere with radius $R$  glued to the end of a microcantilever of linear dimensions $l\times w\times t$. We have employed both the fundamental and first harmonic flexural modes of a soft cantilever (C1), and the fundamental flexural mode of a shorter stiffer cantilever (C2).  Figure \ref{fig_asymptote}(a) depicts optical measurements and finite element method (FEM) simulations of the mechanical modes of C1. For each device and mode, we first extract the the intrinsic quality factor $Q_i$ and resonance frequency ${\omega_i}\over {2\pi}$ in UHV  away from any surfaces.  The modal mass $m_e$ is determined from the resonance frequency shifts before and after the sphere is attached to the cantilever. These parameters are listed in Table \ref{tab_devices}.

Once the mechanical mode is characterized, we change the flow parameters while optically monitoring the dissipation and the resonance frequency of the mode. In particular, we continuously vary two parameters for each mode as follows. i) We change the gap $h$ (shortest distance) between the sphere and a flat solid (Silicon) surface. At small gaps ($h\le200$ nm), we drive the cantilever to achieve `intermittent contact' between the sphere and the solid, and determine the gap from the amplitude. For large gaps, $h$ is extracted from a calibrated linear motion stage. ii) We vary the surrounding pressure $p$ by admitting dry N$_2$ into the chamber. These provide a two-dimensional parametric map of the dimensionless dissipation and the (angular) resonance frequency: ${Q_m}^{ - 1} = {Q_m}^{ - 1}(h,p)$ and  $\omega_m=\omega_m(h,p)$. Before presenting the data, we show in Fig.~\ref{fig_asymptote}(b) that $1/Q_{m}$ measured by linearly driving the resonator and by monitoring its thermal fluctuations agree closely, with a typical discrepancy less than 1$\%$. The maximum  amplitudes in driven and thermal measurements remain $\sim 1$ nm and $\sim 0.01$ nm, respectively. By properly subtracting the intrinsic dissipation from the measured dissipation, one can obtain the fluidic dissipation: $1 / Q_f=1 / Q_{m} - 1 / Q_i $.  Figure \ref{fig_asymptote}(c) and (d) show the ${Q_f}^{ - 1} = {Q_f}^{ - 1}(h,p)$ data set for the 13.7 kHz mode in double-logarithmic plots against gap $h$ and pressure $p$, respectively. In Fig. \ref{fig_asymptote}(c), the gap is varied in the range $10^{-8} $~m $\le h\le 10^{-4} $~m with the pressure held at $p=100, 300$ and 1000 Torr. Conversely, in Fig. \ref{fig_asymptote}(d),  the pressure is swept continuously in the range $ 10^{-2}$~Torr $\le p\le 10^{3} $~Torr, while the gap is fixed at $h=0.1, 0.2, 1$ and 20 $\mu$m. The inset in Fig.~\ref{fig_asymptote}(c) is a semi-logarithmic plot, showing the characteristic saturation of ${Q_f}^{ - 1}$ vs. $h$ (see discussion on $1/Q_{f\infty}$ below). The accompanying mode frequency, $\omega_m=\omega_m(h,p)$, show very little variation (less than $0.1\%$) in this parameter space.

Several important preliminary observations can be made from the data of Fig. \ref{fig_asymptote}(c) and (d).  For  a sphere oscillating at frequency $\omega\over2\pi$ in an unbounded  fluid at the viscous limit $\omega\tau \ll 1$  \cite{Yakhot_Colosqui,Ekinci_Nanofluidics}, the dimensionless dissipation can be written as \cite{Landau}
\begin{equation}
\frac{1}{Q_{f\infty}} = \frac{6 \pi \mu R}{m \omega} \left( 1 + \frac{R}{\delta} \right),
\label{eq_q_infinite}
\end{equation}

\noindent where $R$ is the radius and $m$ is the mass of the sphere, $\mu$ is the dynamic viscosity of the fluid, and $\delta  = \sqrt {{{2{\mu}} \over {\rho \omega }}}  $ is the viscous boundary layer thickness. ${\mu } = {\rho }{\nu }$, where
 $\rho$ is the density and ${\nu }$ is the kinematic viscosity of the fluid. The fluidic dissipation from the rectangular cantilever can also be found, albeit numerically \cite{Sader}. The solid line segments in Fig. \ref{fig_asymptote}(c) and the solid curve in Fig. \ref{fig_asymptote}(d) show the $1 / Q_{f\infty}$ predictions of  viscous theory at large gaps, $h \to \infty $. In these calculations, the independent contributions to dissipation from the sphere and the cantilever are simply added.  The velocity field of an oscillating  sphere-cantilever system should be different from that obtained by adding the individual velocity fields of a sphere and a cantilever. Regardless, the agreement between experiment and calculations in Fig~\ref{fig_asymptote}(c) and (d) is satisfactory \cite{footnote}. In  Fig. \ref{fig_asymptote} (d), the prediction of  molecular theory \cite{Bhiladvala_Molecular} is also shown.

\begin{figure*}
\includegraphics[width=7 in]{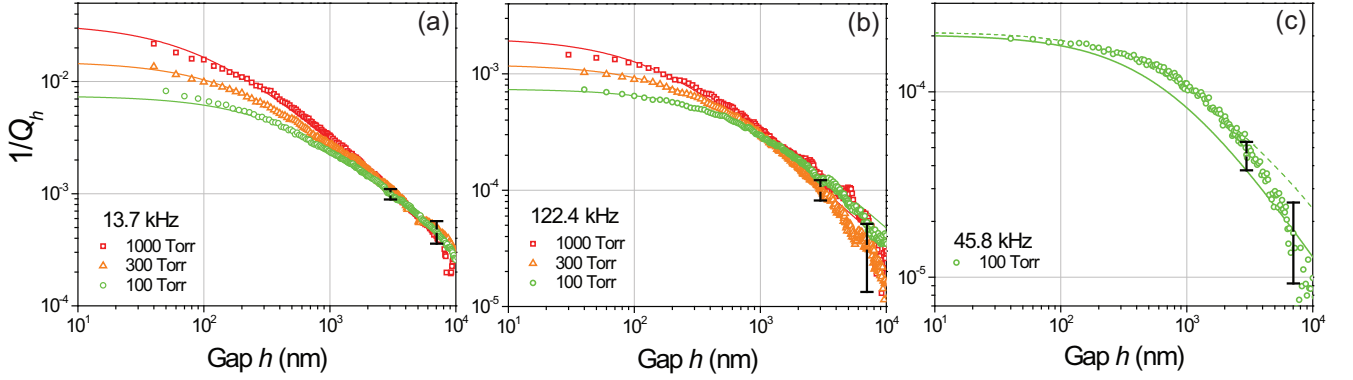}
\caption{\label{fig_fits} Gap dependent dimensionless dissipation $1/Q_h$ as a function of gap $h$ at fixed pressure. (a), (b) Fundamental modes of devices C1 and C2, respectively. (c) First harmonic mode of device C1. Solid lines in (a)-(c) are fits to Eq.~(\ref{eq_q_gap}) with $\alpha=0.5$ and $\beta=1.6$, multiplied by a fitting factor of ${\cal{C}}\approx0.23\pm0.11$.   The deviation from the solid line in (c) is possibly due to the additional rotational motion of the sphere. The dashed line in (c) is the improved fit with the added rotational dissipation \cite{Microhydrodynamics}. The noise in all the data in (a)-(c) increases for $h\ge10^4$ nm  due to the subtraction of $1 / Q_{f\infty}$. The representative error bars are found by an analysis of the noise at the tails ($h\ge10^4$) and become smaller than symbols for $h\lesssim 10^3$ nm.}
\end{figure*}


\begin{table}
\caption{\label{tab_devices}Mechanical properties of the measured devices.}
\begin{ruledtabular}
\begin{tabular}{ccccccc}
Device & Mode &  $l \times w \times t$ & $R$ & $\omega_i \over 2\pi$ &  $Q_i$ & $m_e$\\
& & ($\mu$m) & ($\mu$m) & (kHz) &  & (kg)\\
\hline
C1 & 1 &  $230 \times 40\times 3$ & 35 & 13.7 & 12$\times10^3$ & 5$\times10^{-10}$\\
C1 & 2 &  $230 \times 40 \times 3$ & 35 & 45.8 & 3.4$\times10^3$ & 16$\times10^{-10}$\\
C2 & 1 &  $125 \times 35 \times 4$ & 21.5 & 122.4 & 6.8$\times10^3$ & 1$\times10^{-10}$\\
\end{tabular}
\end{ruledtabular}
\end{table}


When a wall is placed in the proximity of an oscillating sphere, the entire velocity field (not just the field in the gap) will be  modified substantially. Regardless, the dissipation caused by the squeezing of fluid in the gap can be conveniently studied by subtracting the dissipation in an infinite fluid,  i.e., $1 / Q_h = 1 / Q_f - 1 / Q_{f\infty}$. Subtracting the \emph{experimental} $h$-independent $1 / Q_{f\infty}$ asymptotes in Fig.~\ref{fig_asymptote}(c) from the $1 / Q_f$ data results in the dimensionless gap-dependent dissipation $1/Q_h$ in Fig. \ref{fig_fits}(a). Figure \ref{fig_fits} depicts similarly-obtained $1/Q_h$ for three different modes at multiple pressures as the gap is varied.  Solid lines are fits to theory (see below). A first pass analysis of the data can be provided based upon the dimensionless Knudsen number, ${\rm{K}}{{\rm{n}}_h} \equiv {\lambda  \over h}$. When ${\rm{K}}{{\rm{n}}_h}\ll 1$, $1/Q_{h}\propto 1/h$ and can be approximated as \cite{Brenner,Microhydrodynamics,Liao_NIH}:
\begin{equation}
{1 \over {{Q_h}}} = {{6\pi {\mu}R} \over {m\omega }} \times {R \over {{h}}}.
\label{eq_hydrodynamic}
\end{equation}
\noindent At the opposite limit of ${\rm{K}}{{\rm{n}}_h}\gg 1$, the dimensionless dissipation saturates. Between these two limits, there is a well-defined transition from the hydrodynamic to the kinetic regime.

We now provide a theoretical background for the observed transition. Since  $1/Q_{h}\rightarrow 0$ as $h\rightarrow\infty$, we can write a general relation
\begin{equation}
{1 \over {{Q_h}}} = {{6\pi {\mu }R} \over {m\omega }}\times {{R \over h}} \times f\left( {{\lambda  \over h},{\lambda  \over {{\delta }}},{\lambda  \over R},...,{R \over {{\delta }}}} \right).
\label{eq_q_general}
\end{equation}
\noindent The scaling function $f(\{ {x_i}\} )$, which is  analytic in the limit $\{ {x_i}\}\rightarrow 0$,  depends on various dimensionless variables pertaining to different dynamic regimes.  It is clear that the first few $\{ {x_i}\} $ are the familiar Knudsen numbers based on appropriate linear dimensions characterizing  the system: ${\rm{Kn}}_{h}=\frac{\lambda}{h}$, ${\rm{Kn}}_{\delta}=\frac{\lambda}{\delta}$, ${\rm{Kn}}_{R}=\frac{\lambda}{R}$  and so on. The last dimensionless variable, ${R \over \delta } = R\sqrt {{\omega  \over {2\nu }}}  = \sqrt {{{UR} \over \nu }}  = {\rm{R}}{{\rm{e}}_\delta }$, can be regarded as a Reynolds number based on the velocity $U=\omega R/2$. In the limit ${\rm{Kn}}_{i}\rightarrow 0$  and ${\rm{R}}{{\rm{e}}_\delta }\rightarrow 0$,  Taylor expansion gives
\begin{equation}
{1 \over {{Q_h}}} =  {{6\pi {\mu }R} \over {m\omega }}\times {R \over h} \times \left( 1+{{f^{(1)}} + {f^{(2)}} + ...} \right),
\label{eq_q_expansion}
\end{equation}
\noindent where
\begin{equation}
{f^{(1)}} = a_h^{(1)}{\rm{K}}{{\rm{n}}_h} + a_\delta ^{(1)}{\rm{K}}{{\rm{n}}_\delta } + ... + a_{{\mathop{\rm Re}\nolimits} }^{(1)}{\rm{R}}{{\rm{e}}_\delta }
\label{eq_expand_1}
\end{equation}
\begin{eqnarray}
{f^{(2)}} = a_h^{(2)}{\rm{K}}{{\rm{n}}_{\rm{h}}}^{2} + a_\delta ^{(2)}{\rm{K}}{{\rm{n}}_\delta }^{2} + ... + a_{{\mathop{\rm Re}\nolimits} }^{(2)}{\rm{R}}{{\rm{e}}_\delta }^2 \nonumber\\
+ a_{h,\delta }^{(2)}{\rm{K}}{{\rm{n}}_h}{\rm{K}}{{\rm{n}}_\delta } +a_{h,{\mathop{\rm Re}\nolimits} }^{(2)}{\rm{K}}{{\rm{n}}_h}{\rm{R}}{{\rm{e}}_\delta }+ ...
\label{eq_expand_2}
\end{eqnarray}

\noindent  The relative magnitudes of the $a_i^{(n)}$ and the dimensionless parameters $\{ {x_i}\}$  determine the physics of the flow. By varying $\{ {x_i}\}$ over a broad range, one can extract the magnitudes of $a_i^{(n)}$  from experiment.

To gain more insight into the proposed expansion in Eq. (\ref{eq_q_expansion}), let us consider the limits. When $h\rightarrow \infty$ (${\rm{Kn}}_{h}\rightarrow 0$),  dimensionless dissipation due to squeezing disappears, $1/Q_h \rightarrow 0$. This suggests that the first order term in the Taylor expansion in Eq.~(\ref{eq_expand_1}) should not strongly depend on the other Knudsen numbers, ${\rm{Kn}}_{\delta}$, ${\rm{Kn}}_{R}$  and so on. In the limit of small $h$ (${\rm{Kn}}_{h}\gg1$), momentum transfer is dominated by the ballistic impact of the molecules emitted from the stationary plate incident on the moving sphere. The contribution of intermolecular collisions can be neglected. If the thermal molecular velocity  $u_{th}$ is  large, the dimension of the gap $h$ must disappear from the expression for dissipation.

Keeping a finite number of terms in Eq.~(\ref{eq_q_expansion}), one can only hope to find an approximation for the scaling function $f(\{ {x_i}\} )$ valid  in the limit ${x_i}\rightarrow 0$. To obtain an expression valid in the entire range of $\{{x_i}\}$  variation,  one has to keep infinitely many terms. This can be achieved by recasting the scaling function in Eq.~(\ref{eq_q_general}) into a ratio of low-order  polynomials with unknown coefficients to be determined experimentally. The resulting  expression
\begin{equation}
\frac{1}{Q_h} = \frac{6 \pi \mu R}{m \omega} \times \frac{R}{h} \times \frac{1}{1+\alpha\frac{\lambda}{h}\left(1+\beta\frac{R}{\delta}\right)}
\label{eq_q_gap}
\end{equation}

\noindent can be perceived as the simplest Pad\'{e} approximant, which should describe experiments in a broad parameter range. The constants $\alpha$ and $\beta$ are related to $a_i^{(n)}$. It is interesting to note that, in this choice, the term of linear order ${\cal O}({{\mathop{\rm Re}\nolimits} _\delta})$   disappears due to the subtraction, $1/{Q_h} = 1/{Q_f} - 1/{Q_{f\infty }}$. However, the higher order term ${\cal O}($${\rm{K}}{{\rm{n}}_h}{{\mathop{\rm Re}\nolimits} _\delta }$$)$ survives.  In the small-gap limit ${\rm{Kn}}_{h}={\lambda  \over h} \gg 1$, one obtains as prescribed
\begin{equation}
\frac{1}{Q_{h}}\sim \frac{6\pi \rho u_{th} R^{2}}{m\omega\alpha\left(1+\beta \frac{R}{\delta}\right)}.
\label{eq_q_limit}
\end{equation}

Returning to Fig.~\ref{fig_fits}, we now describe how the fits to the experimental data are obtained based upon the above scaling form. The device parameters $m=m_e$, $\omega$, and $R$ are experimental constants. The fluid parameters are all assumed to be independent of $h$, but may depend on $p$:  $\lambda \propto p^{-1}$, $\delta \propto p^{-1/2}$, and $\mu$  is independent of $p$. The very \emph{same} constants $\alpha$ and $\beta$  in the scaling function in Eq.~(\ref{eq_q_gap}) must uniquely fit \emph{all} data sets --- regardless of pressure, frequency, mode and so on. Indeed, we can fit \emph{all} our data with $\alpha=0.5$ and $\beta=1.6$, found by iteration. Any small changes in $\alpha$ and $\beta$  cause the curves in Fig.~\ref{fig_fits} to shift along the $h$-axis, making the fits unacceptable. The fits can be improved along the $1/Q_h$-axis by multiplying with fitting factors of ${\cal{C}}\approx0.33,~0.20$ and 0.16 for the 13.7 kHz, 122.4 kHz and the 45.8 kHz modes, respectively, resulting in the solid curves in Fig.~\ref{fig_fits}. To within our experimental accuracy, however, ${\cal{C}}$ remains a constant as ${\cal{C}}\approx0.23\pm0.11$ for all our devices, and  may be needed due to  non-idealities in geometry (e.g., the cantilever and epoxy above the sphere), inaccuracies in determining $m_e$  (especially for the first harmonic mode) and deviations  from normal relative motion (see below). Deeper physical factors --- such as the non-trivial effects of the subtraction of the $1/Q_{f\infty}$ tails and unsteady corrections to Eq.~(\ref{eq_hydrodynamic})--- cannot be ruled out, and may give rise to the small deviations in $\cal{C}$ from device to device.

The fit in Fig. \ref{fig_fits}(c) (solid curve)  deviates from the data for $10^2$~nm $\lesssim h \lesssim 5\times10^4$~nm. FEM simulations for this mode suggest that the sphere undergoes rotational motion --- with the displacement of its closest point to the wall being in the direction $0.98{\bf{\hat x}}+0.2{\bf{\hat z}}$ [see Fig.~\ref{fig_asymptote}(a)]. Then, the dissipation comes from shearing the fluid in the gap as well as from squeezing it. For shear, ${1 \over {{Q_h}}} = {{48\pi \mu R} \over {15m\omega }} \times \ln \left( {{R \over h}} \right)$ \cite{Microhydrodynamics} as opposed to the expression in Eq.~(\ref{eq_hydrodynamic}) for squeezing.  The dashed line in Fig. \ref{fig_fits}(c) is the fit found by na\"\i vely adding these two forms in the ratio of the FEM motional amplitudes, and by keeping the scaling function exactly the same. Because the effect remains small, the  $1/h$ dependence of the dissipation can be assumed prevalent for all devices considered here and in the literature.


\begin{figure}
\includegraphics[width=3.375in]{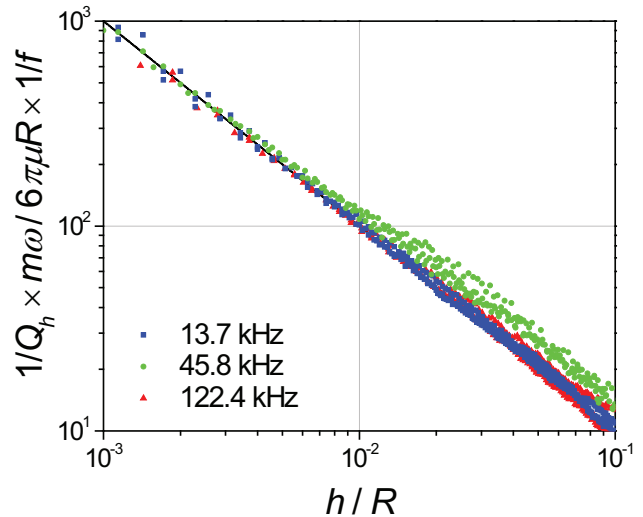}
\caption{\label{fig_universal}  Collapse of all experimental data from this work. Here, $f$ is the scaling function defined in Eq.~(\ref{eq_q_gap}) and the same $\cal{C}$s are used as above.}
\end{figure}


Having fit individual  data traces, we can  collapse all our data as shown in Fig.~\ref{fig_universal}. The collapse is obtained by removing the trivial effects of the device size and frequency from the data as well as the more profound  effects of the scaling function $f(\{ {x_i}\} )$. The plotted dimensionless quantity, ${1 \over {{Q_h}}} \times {{m\omega } \over {6\pi \mu R}} \times {1 \over f}$ can be regarded as the dimensionless size- and frequency-independent dissipation, in which the kinetic effects have been deconvoluted. It therefore shows the hydrodynamic $R/h$ dependence at all length scales studied here.


Finally, our results can be interpreted as follows.  In the hydrodynamic limit  ($h\gg\lambda$), this problem is described by Eq.~(\ref{eq_hydrodynamic}), where the viscosity $\mu$ is dominated by intermolecular collisions, $\mu \sim \rho u_{th}\lambda$, with a relaxation timescale $\sim\lambda/u_{th}$.  To gain insight into the kinetic limit ($h\ll \lambda$), one can simply write the shear stress on the sphere as $\sigma \sim \rho u_{th} \left| {\dot h} \right|$. It is easy to see that  $\sigma \sim \rho u_{th} h \frac{\left| {\dot h} \right|}{h}\sim \rho u_{th} h \frac{du}{dz}$, where $\frac{du}{dz}$ is the velocity gradient.  This result can be interpreted as the appearance of an effective viscosity, $\mu_{eff}\approx \rho u_{th} h$, due to an effective mean free path, $\lambda_{eff}\approx h$. Substituting $\mu_{eff}$ into the hydrodynamic solution simply results in ${1 \over {{Q_h}}} \sim {{ \rho {u_{th}}{R^2}} \over {m\omega }}$, consistent with Eq.~(\ref{eq_q_limit}). Thus, in principle, one may justify an attempt to reach the kinetic regime by using the Navier-Stokes equations, but combined with effective (and sometimes  frequency-dependent) viscosities, slip lengths and so on.

In this manuscript, we have presented experimental data on confined nanoflows covering a broad range of flow parameters. Our simple scaling theory  describes experiments in the entire parameter range --- without explicitly employing an effective viscosity and/or slip length.  To conclude, we stress that the dimensionless Weissenberg numbers here remain small,  ${\rm{Wi}} = \omega \tau  \ll 1$. Since  the appearance of frequency in effective viscosity essentially leads to a modification of the equations of motion \cite{Ekinci_Nanofluidics, Yakhot_Colosqui}, generalization of confined nanoflows to the interval $\rm{Wi}\gg1$ will require further experimental and theoretical work.


\begin{acknowledgments}
The authors  acknowledge generous support from the US NSF (through grants
DGE-0741448, ECCS-0643178 and CMMI-0970071).
\end{acknowledgments}



\begin{thebibliography}{99}
\bibitem{footnote} For the 13.7 kHz and 122 kHz modes, the ratio of the measured and calculated   $1/Q_{f\infty}$  agree closely, with an error $\lesssim 15 \%$. For the 45.8 kHz mode, the measured $1/Q_{f\infty}$ is a factor of 4 larger than the calculated value. This is possibly due to the more complex motion of the sphere.
\end{thebibliography}
\end{document}